\documentclass[journal,comsoc]{IEEEtran}

\usepackage[T1]{fontenc}% optional T1 font encoding

\ifCLASSINFOpdf
   \usepackage[pdftex]{graphicx}

  \DeclareGraphicsExtensions{.pdf,.jpeg,.png}
\else

\fi

\usepackage{amsmath}

\usepackage[cmintegrals]{newtxmath}

\usepackage[caption=false]{subfig}
\usepackage{algorithm}
\usepackage[noend]{algpseudocode}

\usepackage{graphicx}
\usepackage{latexsym}
\usepackage{amsmath,epsfig}
\usepackage{epstopdf}
\usepackage{caption}
\usepackage{multirow}
\usepackage{booktabs,siunitx}
\usepackage{lscape}
\usepackage{url}
\usepackage{graphicx}
\usepackage{epsfig}
\usepackage{graphicx}
\usepackage{latexsym}
\usepackage{amsmath}
\usepackage{amssymb}
\usepackage{mhsetup}
\usepackage{mathtools}
\usepackage{mathrsfs}
\usepackage{float}
\usepackage{xfrac}
\usepackage{epsfig}
\usepackage[english]{babel}
\usepackage{subfig}
\usepackage{thmtools}
\usepackage{thm-restate}
\usepackage{cleveref}

\newtheorem{thm}{Theorem}
\newtheorem{lem}[thm]{Lemma}

\begin{document}

\title{An Energy Efficient Distributed Gossip Algorithm for Wireless Sensor Networks based on a Randomized Markovian Duty-Cycling}

\author{Ramadan~Abdul-Rashid \\ 
	
Electrical Engineering Department, King Fahd University of Petroleum and Minerals, Dhahran, Saudi Arabia \\
Email: {ramadan.rashid.bukari@gmail.com}
}

%\thanks{Manuscript received on; revised }}

\maketitle

\begin{abstract}
This paper proposes a  novel asynchronous consensus algorithm which is based on a continuous update rule and an energy efficient event triggered duty (wake-sleep) cycle based on a discrete Markov chain model. The system model of the proposed algorithm is formulated and analyzed. The conditions for convergence and stability of the algorithm are derived and the algorithm is proved to converge to an average consensus. Numerical simulations on random, circular, chain and star graphs show stability in consensus and convergence to network global state average. 
\end{abstract}

\IEEEpeerreviewmaketitle

\section{Introduction}
A set of nodes in a distributed network can communicate through the exchange of information where each node initially holds a measured or computed parameter and wants to learn in a distributive way, the average of all the measurements of the other nodes in the network. Distributed average consensus algorithms are designed to solve this problem. Node units in the network do not necessarily have a thorough global knowledge about the network. For example nodes might not be aware of the number of nodes, the network topology, or the type of quantities collected at other nodes, etc. Moreover, in some applications or network frameworks, the network topology can vary with time due to link instability or node mobility. The goal of the average consensus algorithm is to reach consensus in a reliable and robust manner.
Average consensus algorithms operate iteratively and the instantaneous value at each node is an estimate of the measurements' average. These algorithms are designed such that all the estimates in a particular network converge to the sought average up to any desired level of precision. The iterative process is classed into three parts.

\begin{enumerate}
	\item First one or several nodes wake up.
	\item Then the woken nodes send their estimates to one or several neighbors in the network.
	\item Finally each receiving node updates its estimate to a value which depends on its current estimate and on the estimates it has received from the woken nodes.
\end{enumerate}

In a synchronous algorithm, all the nodes in the network wake up at each instance of iteration and broadcast their estimates. All nodes in the network then update their estimates before the next iteration instance can begin. 
On the other hand, in an asynchronous algorithm, only one node or a subset of nodes wake up at each iteration. These node(s) call some chosen neighbors. Only a subset of nodes in the network update their estimates at the end of each iteration. This work follows the asynchronous approach to consensus. Asynchronous consensus or gossip algorithms have received quite a fair attention in several areas such as distributed systems, network optimization, robotics and multi-agent wireless systems. Traditional gossip algorithms include, the pairwise gossip \cite{boyd_randomized_2006}, neighborhood gossip \cite{avrachenkov_local_2011}, geographic gossip and path averaging \cite{denantes_which_2008} algorithms. Most recent algorithms are variants of these conventional algorithms \cite{abdul2018accurate}, where authors have focused on optimizing algorithm performance in terms of energy utilization and convergence time where node with a randomized duty cycling of nodes.

In a WSN, each network node always possesses its initial measurement(s), with no knowledge of the measurement of other nodes. To relay information about the state of a sensor field, the WSN is required to transmit each node's measurements to a certain control station at each instant of measurement, which requires frequent communication, which in-turn depletes node energy and hence network lifetime \cite{8893220}. To address this problem, average consensus and gossip or asynchronous distributed algorithms are designed to compute neighborhood averages of node measurements iteratively until converging to a consensus, which is transmitted as the state of the sensor field. In this research, we propose a novel improved neighborhood gossip algorithm, which models the neighborhood of each node as a subsystem and uses a discrete switching mechanism to select nodes to compute neighborhood measurements in an asynchronous manner until a global network consensus is reached. The performance of the method is evaluated using both numerical simulations on different network topologies. 

The key contributions of this paper are outlined below.
\begin{enumerate}
	\item Proposal of a novel method for gossip algorithm for wireless sensor networks
	\item Discuss the stability and convergence analysis of the proposed algorithm
	\item Experimental numerical evaluation on the performance of our suggested algorithm.
\end{enumerate}

The rest of the paper is organized as follows. Section II discusses the system model. Section III presents the generalized gossip algorithm and section IV presents our proposed algorithm. Convergence analysis of the algorithm is presented in section VI. Numerical comparative evaluation of the proposed gossip algorithm is then presented in Section VII. The paper concludes in Section VII.

\section{System Model}
In a sparsely distributed Wireless Sensor Network (WSN), ordinary nodes spread across a relatively localized area or sensor field periodically report their measurements' or data to an anchor or gateway node \cite{abdul2018time}. Unlike the ordinary nodes with limited energy, bandwidth and memory, most gateway nodes have a bulk of these resources. Additionally in many wireless sensor network architectures, the gateway node has access to a stable and precise clock reference, for example a GPS receiver \cite{sommer_gradient_2009,akl_investigation_2011}. These intermediate or anchor nodes normally broadcast beacons to nodes within their neighborhood for some network establishment or management services as shown in Figure \ref{WSN}. The gossip algorithm considered in this work utilizes this beaconing to regulate the wakeup of ordinary nodes so as to minimize collisions and redundant updates which depletes node energy.

\begin{figure}[htbp!] 
	\centering    
	\includegraphics[width=0.5\textwidth]{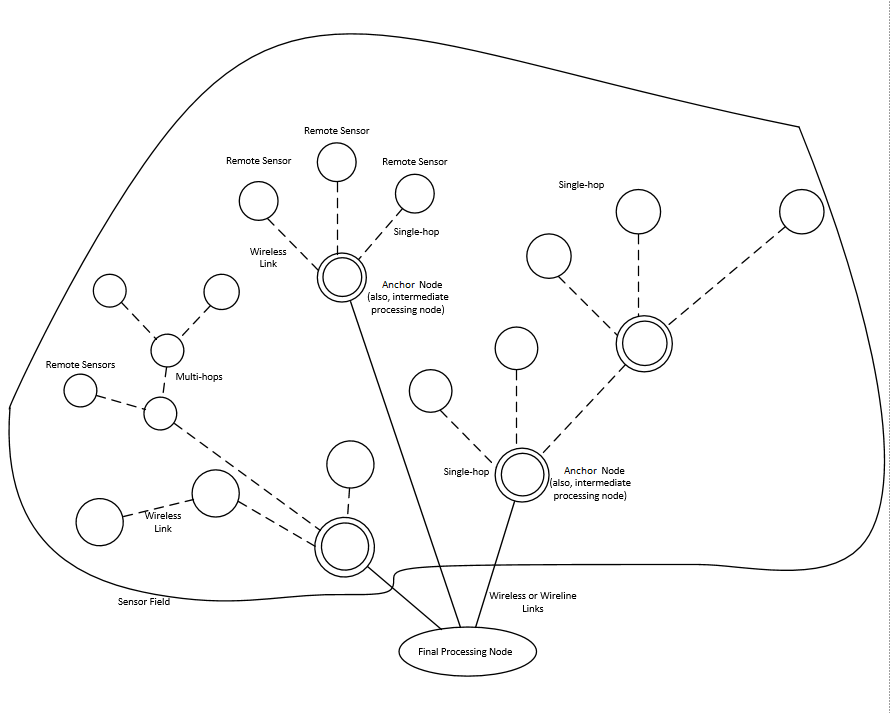}
	\caption[Typical Sensor Network Arrangement]{Typical Sensor Network Arrangement}
	\label{fig:WSN}
\end{figure}

%\subsection{Network Model}
Assume a WSN has symmetric links and hence can be represented by an undirected graph $\textbf{g}=(\textbf{v},\textbf{e})$. In this model, we represent the sensor nodes of the network as the vertex set, $\textbf{v}=\{i|i = 1,2,\hdots ,N\}$, where $N = |\textbf{v}|$ is the cardinality of $\textbf{v}$ and the working network connectivity between these nodes as an edge set, $\textbf{e} \ \text{such that} (i,j) \in \textbf{e}$ if nodes $i$ and $j$ can send information to each other. Such nodes that directly communicate with node $i$ are referred to as the neighborhood nodes of $i$ and represented by the set, $n_i = \{j|\textbf{e}_{i,j}\in \textbf{e}\} $.
 
\section{Generalized Gossip Algorithm}
A gossip algorithm is a distributed iterative algorithm, where at each iteration $k$, a random subset $s(k)$ of nodes update their estimates to the average of the estimates of $ s(k)$: for all $j \in s(k)$

\begin{equation}\label{eq_main}
x_i(k) =   \frac{1}{|s(k)|} \sum_{j \in s(k)}^{|n_{i}|}x_j(k-1) 
\end{equation}

In both standard gossip and geographic gossip, $s(k)$ always has two nodes, whereas in neighborhood gossip and in path averaging, where whole neighborhoods and paths are being averaged at each iteration, $s(k)$ has a random number of nodes.

\subsection{Distributed Gossip Algorithms as Switched Systems}
In our conception of distributed consensus algorithms, we consider the facts that, in sparsely or randomly distributed WSN, there exist a high chance that, if ordinary node positions are partitioned based on 1-hop communication to the anchor node, nodes lying within nearly the same range as the anchor node at any instant $k$ communicate with the gateway node as the same time. Utilizing this feature, we design a wakeup protocol, regulated by the anchor node, where all nodes in the same \textit{synchronous layer} or within similar distances to the anchor node, carry out state updates at the same time. Hence we formulate the state update of a node $i$ as: 

\begin{equation}\label{eq_update}
x_i(k) =   \frac{\phi_{i}(k)}{|n_{i}|} \left [\sum_{j \in n_i}^{}x_j(k-1) \right ]
\end{equation}

where,
$$
\phi_{i} =
\begin{cases}
1, & \text{if}\ \text{node}\ i\ \text{updates}\  \text{at}\ \text{time} \ k\\
0, & \text{otherwise}
\end{cases}
$$

Hence at time instant, $k$ a generalized update equation can be written as

\begin{equation}\label{eq_asyn}
\left[ \begin{array}{c} \textbf{x}_{k} \\ \Phi_k \textbf{1}\end{array} \right] = \left[ \begin{array}{c} \Phi_k \textbf{A}\textbf{x}_{k-1}  \\ \mathcal{F}(\Phi_{k-1} \textbf{1})\end{array} \right]
\end{equation}

where
$$
\Phi_{k} = \begin{bmatrix}
\phi_{1}(k) &  0  & \ldots & 0\\
0  &  \phi_{2}(k) & \ldots & 0\\
\vdots & \vdots & \ddots & \vdots\\
0 &  0  & \ldots & \phi_{N}(k)
\end{bmatrix}
$$

, $\textbf{x}_k$ which is a vector of the entire network state estimates also expressed as: $\textbf{x}_k = [x_1(k)\  x_2(k)\  x_3(k)\  \cdots\  x_{N}(k)]^{\textbf{T}}$ and \textbf{1} is a vector of ones.

From \ref{eq_asyn}, we can describe a generalized gossip algorithm as a dynamic system with a finite number of subsystems represented by the nodes updating at $k$ and a logical rule that triggers switching between these subsystems represented by the wake-up protocol used to wake op nodes lying within the same synchronous layer. This framework of asynchronous update is reminiscent of a linear switched control system \cite{abdul2018accurate}.

\section{Proposed Gossip Algorithm}
\begin{figure}[ht!] 
	\centering    
	\includegraphics[width=0.450\textwidth]{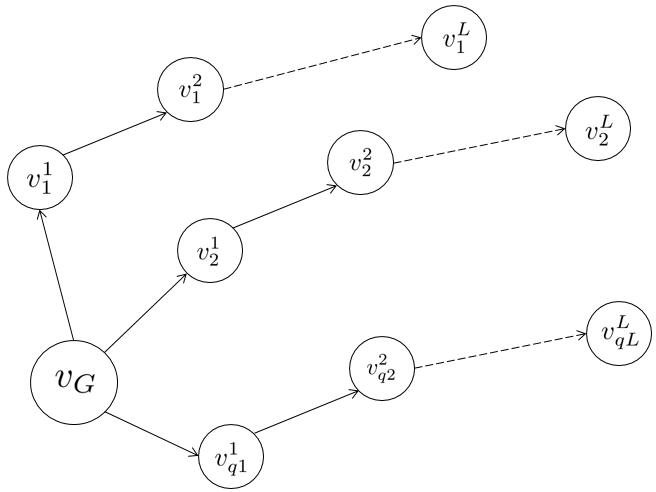}
	\caption[Network Partitioning into Synchronous Layers]{Network Partitioning into Synchronous Layers}
	\label{fig:net_1}
\end{figure}

\begin{figure}[ht!] 
	\centering    
	\includegraphics[width=0.450\textwidth]{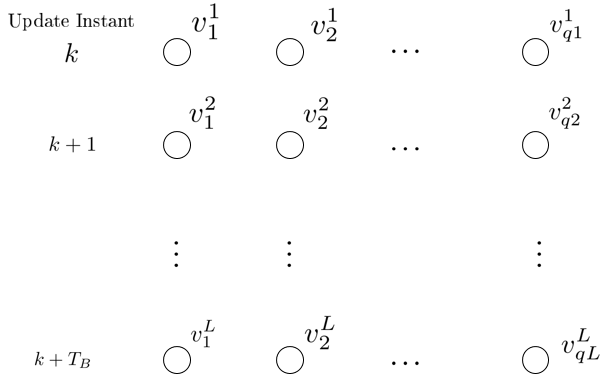}
	\caption[Node Update Times and Wake-up Schedule]{Node Update Times and Wake-up Schedule}
	\label{fig:net_2}
\end{figure}

Consider the described network model, let $v_G$ be the anchor and number of ordinary nodes, $N$. Assume the network has $L$ layers with $qm$ nodes per layer. Let node $v^{m}_n$ be the $n^{th}$ node in the $m^{th}$ synchronous layer, where $n = 1, \hdots, qm \ \text{and} \ m = 1, \hdots, L$. A conceptual partition of a network into $L$ synchronous layers based on proximity to anchor node is given by Figure \ref{fig:net_1} and the wake-up update schedule for each node for the first cycle of update is given by Figure \ref{fig:net_2}.\\

\noindent
First we define the following parameters. Let:

$d$ be 1-hop communication delay, $d \sim \mathcal{N}(0, \sigma^{2}_{d})$ 

$t_C$ be the time needed to receive and compute average 

$t_W$ be the time between sleep and wake-up\\ 

\noindent
And the following events:

Transition of $\phi_{i}$ from $1 \longrightarrow 0$  at time $t_C$ 

Transition of $\phi_{i}$ from $0 \longrightarrow 1$  at time $t_W = (L-m)(d + t_C)$ 

$v_G$ broadcasts a beacon packet every $T_s  = L (d + t_C) \sigma^{2}_{d}$\\

Without loss of generality, an arbitrary network node, $v_i$ is referred to as node $i$ for the rest of our presentation. We outline the stages of operation of our proposed algorithm in the pseudo-code \textbf{Algorithm} \ref{alg1} which are described as follows:

\begin{enumerate}
	\item Each node, $i$ has a binary status variable $\phi_i$ that is set to, $\phi_i=0$. Let us assume an upper bound $L$ on the number of connectivity layers, where $L$ depends on the network size and topology. 
	
	\item The gateway node initializes update timer $T_s$ and triggers the update of the nodes connected to it.
	\item Once a node $i$ updates, it triggers the update of its nearest neighbor nodes, $j$ whose status bit variable, $\phi_j$ are a complement of its own, i.e., $\phi_i=\acute{\phi}_{j}$. Once the flooding of the status bits variable begin, if $i$ receives  $<\phi_j = 1>$ or when a beacon node is received from the anchor node, node $i$ awakes and switches to active mode. 
	
	\item Once node $i$ awakes, it requests for states values from its neighbors (in 1-hop), and upon receiving state values from $n_i$ nodes, it computes the current estimate from the received states. 
	
	\item Once the update is computed, node $i$ sets $\phi$ to 1, broadcast a wake-up and goes back to inactive mode. It then sets its $\phi$ to 0 after switching to inactive mode. 
	
	\item Node $i$ however listens for wake-up signal and state request packets. When a state request packet is received from another node $l$ which lies within 1-hop to node $i$, i.e., $l \in n_i$, node $i$ replies with an acknowledgment with $x_i$ as payload.
	
	\item This process continues until the timer of the anchor node is $T_s$ seconds. When this event is true, the anchor node initializes its update time $T_s$ and trigger the update of the nearest nodes and hence the whole process begins again.
\end{enumerate}

\begin{algorithm}
	\caption{Pseudo-code for Node $i$}\label{alg1}
	\begin{algorithmic}[1]
		\State For any node $i$, let $x_i(0)$ be its measurement, and let $K$ be a number of iterations.
		\State  \ $\phi_i \gets 0$; \Comment{Binary status activation parameter}
		\If{ $ \phi_j $ is received from $j$ such that $\phi_{j} = 0$ OR Beacon packet is received from anchor node}		
		\State Node $i$ wake up to update
		\Else{\ Node $i$ remains in inactive mode} \Comment{conserves energy}
		\EndIf 
		\For{$k = 1:K$}
		\State Node $i$ request state estimates from $n_i$ nodes
		\State Say node $j$, is such that $\{j\in n_i| n_{i} \in \textbf{v} \ \text{and} \ \textbf{e}_{i,j} \in \textbf{e}\}$ 
		\If{<<$x_j$>> request is received at node $j$} 
		\State $j$ sends an acknowledgment with payload <<$x_j$>> 
		\EndIf

		\State Let $n_i$ be the set of nodes that send state estimates to \\
		 \ \ \ \ node $i$ at time instant $k$. Node $i$ computes
		\State $$x_i(k+1) \gets \frac{1}{|n_i|}\sum_{j \in n_i}^{} x_j(k-1) + x_i(k-1)$$
		\State Set $1 \gets \phi_i $ and broadcast $\phi_i$ wake-up message to $n_i$ \\
		 \ \ \ \ nodes in next layer 
		\State Node $i$ switches to inactive mode \Comment{conserves energy}
		\State \textbf{Upon receiving <<$x_i$>> estimates request from node $l$}\\ \Comment{Given that $i \in n_l$}		
		\State Node $i$ transmits acknowledgment <<$x_i$>> to node $l$
%		\State \textbf{Upon every $t_W$ seconds}
%		\State Node $i$ wakes up to reinitialize the algorithm
		\EndFor
	\end{algorithmic}
\end{algorithm}

\subsection{Event Triggered Wake-Up Model}
\begin{figure}[ht!] 
	\centering    
	\includegraphics[width=0.450\textwidth]{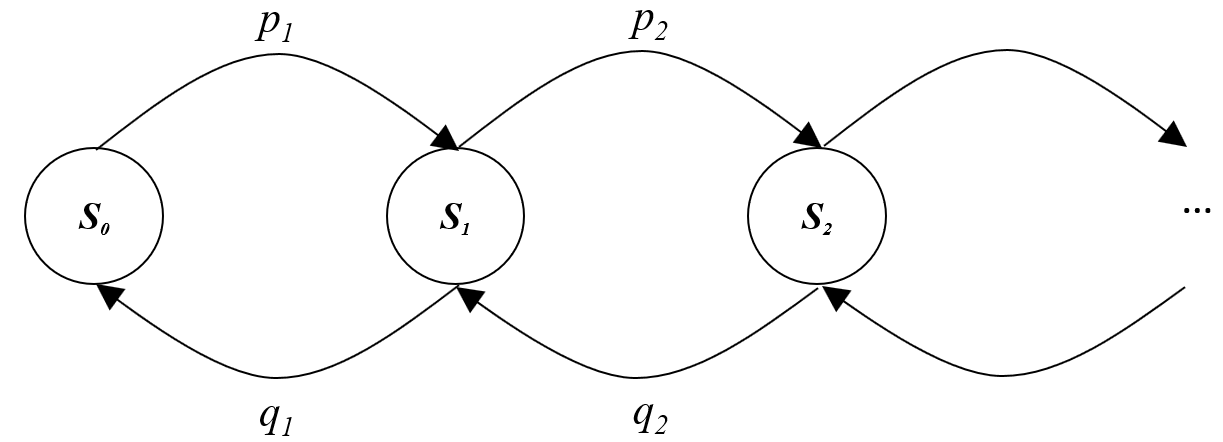}
	\caption[Event Triggered Wake-up Model as a Markov Chain]{Event Triggered Wake-up Model as a Markov Chain (\textit{Random Walk})}
	\label{fig:markov}
\end{figure}

Where node $i$ is activated for update when $\phi_i(k)$ is $'1'$ and inhibited from update when $\phi_i(k)$ is $'0'$. The activation variables are used to control the state update of each node and is determined by the graph of the wireless sensor network. The subset of nodes that update at any time instant also depend on their respective virtual clock values in the previous instant. Based on the wake-up protocol, the event triggered wake-up activation behavior is reminiscent to death-birth process with two binary states $'0'$ and $'1'$ as shown in Figure \ref{fig:markov}. \\

\begin{lem}\label{T1}

	Let $z = [z(k) : k \geq 1]$ be a sequence of iid $\mathrm{R}^{d}$-valued random variables. Consider the sequence $\phi = [\phi(k) : k \geq 0]$ defined through the recursion:
	\begin{equation}\label{bzz}
	   \phi_i(k) = \phi_i(k-1) + z(k)
	\end{equation}
	
	where $\phi_i(0)$ is independent of $z$ and $z(k)$ is updated based on (\ref{azz}), with $d=1$; 
	\begin{equation}\label{azz}
	    z_i(k+n) =
        \begin{cases}
        +1, & n = 1,3,5, \hdots \\
        -1, & n = 2,4,6, \hdots
    \end{cases}
	\end{equation}
	
	This recursion represents a special Markov chain called the \textit{random walk}.
	The system presented by (\ref{bzz}) and (\ref{azz}) can be used to represent the event triggered wake-up model presented in Algorithm \ref{alg1}.

\end{lem}

Based on Lemma \ref{T1} the dynamics of the activation parameter $\phi_i$ for node $i$ is given by (\ref{eq_act}) and generalized for the whole network in (\ref{eq_act1}). 

\begin{equation}\label{eq_act}
\phi_i(k) = \phi_i(k-1) + z_i(k)
\end{equation}
\begin{equation}\label{eq_act1}
\Phi_{k}\textbf{1}  = \Phi_{k-1}\textbf{1} + Z_k
\end{equation}

Assuming the transition between an inactive state $\phi_i(k) = 0$ to and active state $\phi_i(k) = 1$ occurs with a probability, $p_{i}$ occurs within $t_W$ seconds and the probability of the converse event occurring denoted as $q_{i}$ occurs within the time $t_C$ seconds, i.e., $P[z_i(k) = 1] = p_i = 1 - P[z_i(k) = -1] = q_i$. From the birth-death Markov process, these probabilities can be related by:
\begin{equation}
p_{i} t_W = t_C q_i
\end{equation}

\subsection{Proposed Algorithm Model}
From Algorithm \ref{alg1}, we can write an update equation for a node $i$ can be given by:

\begin{equation}\label{eq7}
x_i(k) = \frac{\phi_{i}(k)}{|n_i|}\sum_{j=1}^{|n_i|} x_j(k-1) + x_i(k-1)
\end{equation}

Based on (\ref{eq7}) we derive a state space discrete linear equations for the nodes states values and activation equation given by (\ref{eq8}) and (\ref{eq_act1}). The entire asynchronous (gossip) average consensus algorithm based on Markovian switching is then given by (\ref{eq21}). 

\begin{equation}\label{eq8}
\textbf{x}_{k} = \Phi_k \textbf{A} \textbf{x}_{k-1}
\end{equation}

%\begin{equation}\label{eq9}
%\textbf{x}_{k} = \Phi_k \textbf{A} \textbf{x}_{k-1}
%\end{equation}

\begin{equation}\label{eq21}
\left[ \begin{array}{c} \textbf{x}_{k} \\ \Phi_{k}\textbf{1} \end{array} \right] =   \begin{bmatrix} \Phi_k \textbf{A} & 0 \\  0 &  \textbf{I}  \end{bmatrix} \left[ \begin{array}{c} \textbf{x}_{k} \\ \Phi_{k}\textbf{1}  \end{array} \right] +   \left[ \begin{array}{c} 0 \\ Z_k \end{array} \right] 
\end{equation} 

where \textbf{A} is the connectivity matrix. 

Equation (\ref{eq21}) can be simplified as:
\begin{equation}\label{eq22}
    \textbf{Y}_k = \textbf{W}_k \textbf{Y}_{k-1} + \textbf{D}_k
\end{equation}
And has a solution,
\begin{equation}\label{eq23}
    \textbf{Y}_k = \textbf{W}^k\textbf{Y}_{0} + \sum_{j=0}^{k-1}\textbf{W}^{k-j-1}\textbf{D}_k
\end{equation}

where:\\ 
the weight matrix of (\ref{eq21}) is:
$$
\begin{array}{c} \textbf{W}(k) \end{array}  =   \begin{bmatrix} \Phi_k \textbf{A} & 0 \\  0 &   \textbf{I}  \end{bmatrix},
$$

$$\textbf{Y}_k = \left[ \begin{array}{c} \textbf{x}_{k} \\ \Phi_{k}\textbf{1} \end{array} \right], \ \text{and} $$ 

$$ \textbf{D}_k = \left[ \begin{array}{c} 0 \\ Z_k \end{array} \right]$$

Let $\textbf{x}(0)$ be the vector of measurements, $x_{avg}$ be the average of $\textbf{x}(0)$, and $\textbf{x}(k)$
be the vector of estimates at time $k$. The vector of all ones is denoted by \textbf{1}.
The convergence question in distributed averaging algorithms is twofold:
\begin{enumerate}
	\item   Do the estimates converge to a consensus? In other words,
is there a scalar c such that

\begin{equation}\label{cons}
\textbf{x}(k) = c\textbf{1}?
\end{equation}

 \item If there is such a consensus scalar $c$, is it equal to the average $x_{avg}$?

\end{enumerate}

\section{Stability and Convergence of Proposed Algorithm}
\subsection{Convergence}
To address the first question of consensus, we look at the solution of (\ref{eq21}). According to (\ref{eq22}) consensus of the form (\ref{cons}) will be achieved if $\lim_{k\to\infty} \textbf{W}_k$ exists. For this to be verified, a general rule for asymptotic stability and convergence in consensus will be if the spectral radius $\rho$ of the weight matrix, $\textbf{W}$ is such that \cite{boyd_randomized_2006}:
\begin{equation}
\rho(\textbf{W}(k)) = \max_{i} |\lambda_i(\textbf{W}(k))| \leq 1, \
i = 1, 2, \hdots, N. 
\end{equation}
$$ \text{where} \ N = |\textbf{v}| = \sum_{l=1}^{L} q_{l}$$

\begin{thm}\label{T1}
	An asynchronous average consensus algorithm of the form described in Algorithm 1, operating on a graph, $\textbf{g} = (\textbf{v},\textbf{e})$ and represented by the state equation (\ref{eq21}) will drive the state measurements of nodes $ \textbf{x}_k$ in a WSN to consensus. 
\end{thm}

%\begin{proof}{\textit{Proof}:}

\textit{Proof.} From (\ref{eq21}) we observe that, the lower \textit{modal} block has constant eigenvalues of $\lambda_i(\textbf{I}) = 1$, which is a stable mode. Therefore the network achieves stable consensus if and only if:
\begin{equation}
\rho(\textbf{A}_\Phi) \leq 1 
\end{equation}
\noindent
where $\textbf{A}_\Phi = \Phi_{k} \textbf{A}$
Assuming we can obtain the SVD decomposition of $\textbf{A}_\Phi$ given by $$\textbf{A}_\Phi = \textbf{Q}\Lambda \textbf{V}$$, where $\textbf{V} = \textbf{Q}^{-1}$. This can also be represented by:
\begin{equation}
    \textbf{A}^{k}_\Phi = \sum_{i=1}^{N}\lambda_i^k(\Phi_{k} \textbf{A})\textbf{q}_i \textbf{v}_i^{T}
\end{equation}
\noindent
where $\textbf{Q} = [\textbf{q}_1, \textbf{q}_2, \hdots, \textbf{q}_N]$ and $\textbf{V} = [\textbf{v}_1, \textbf{v}_2, \hdots, \textbf{v}_N]$

$\textbf{x}(k)$ converges asymptotically to $c = \frac{1}{N} \textbf{1}^{T} \textbf{x}(0)$ if and only if [\cite{pereira2011distributed}]:
\begin{equation}\label{eq24}
\textbf{A}_\Phi \textbf{1} = \textbf{1} \  \text{, and} \ 
\end{equation}
\begin{equation}\label{eq25}
    \rho(\textbf{A}_\Phi - \textbf{1}\textbf{v}^{T}_{1}) < 1
\end{equation}
\noindent
where $\rho(\textbf{A}_\Phi - \textbf{1}\textbf{v}^{T}_{1})$ is the second largest eigenvalue associated with $\textbf{A}_{\Phi}$.\\

For (\ref{eq24}) first note that the connectivity matrix, $\textbf{A}$ is such that $\textbf{A}_\Phi = 1$ for average consensus algorithm. This is because for average synchronous average consensus, the row-sums of the connectivity matrix $\textbf{A}$ is always 1.  Further, the state transition matrix, $\Phi$ is such that,
If only one node in the network updates at a time as in our algorithm, then one element in the diagonals of $\Phi$ is 1 say in an arbitrary row $m$. Hence the same elements in row $m$ of $\textbf{A}$ will be retained in $\Phi \textbf{A}$. Therefore all row-sums of $\Phi \textbf{A}$ will be zero except for row, $m$ which will be one.

Furthermore, to prove (\ref{eq25}), we decompose the matrix, $\Phi \textbf{A}$. For our algorithm where only one node, $i$ updates at a time, $k$, the diagonal matrix, $\Lambda^{M\times N}$ of eigenvalues has entries;
$$
\lambda_{m,n}(k) =
    \begin{cases}
    \phi_{i} \lambda_i(\textbf{A}), & m = n = i \\
    1, & m = n \neq i \\
    0, & \text{elsewhere}
\end{cases}
$$
Hence at time $k$, when only node $i$ is updating, the second largest eigenvalue is always zero, i.e., 
$$\rho(\textbf{A}_\Phi - \textbf{1}\textbf{v}^{T}_{1}) = 0$$ This statement also hold for all nodes in the graph, $\textbf{g}$

This proves theorem 2.
%\end{proof}

\subsection{Convergence to Average Consensus}
From \cite{denantes2008distributed}, the conditions for convergence of our distributed consensus algorithm to the average, $\textbf{x}_{avg}$ can be given as:
\begin{equation}\label{eq26}
   \textbf{1}^{T}\textbf{A}_{\Phi} = \textbf{1}^{T} 
\end{equation}
\begin{equation}\label{eq27}
    \rho(\textbf{A}_{\Phi} - \textbf{J}) < 1
\end{equation}
\noindent
where $\textbf{J} = \textbf{1}\textbf{1}^{T} / N$

Due to randomized nature of the event triggered duty-cycling of nodes in this algorithm, the matrix $\textbf{A}_\Phi$ is time variant, i.e. $\textbf{A}_\Phi = \textbf{A}_\Phi (k)$. Due to the time variant nature of $\textbf{A}_\Phi$, conditions (\ref{eq26}) and (\ref{eq27}) might not hold. In \cite{benezit2009distributed}, the weight matrix, $\textbf{W}$ presented for pairwise-gossip \cite{denantes2008distributed}, has entries given by (\ref{eq28}) which is similar to the generalized structure of $\textbf{A}_\Phi$, except $\textbf{A}_{\Phi_{ji}} = 1 \neq \textbf{A}_{\Phi_{ji}} = \beta$ if and only if $j$ is in the
neighborhood, $n_i$ of node i, including node i itself, and $\textbf{A}_{\Phi_{ii}}=0 \neq \textbf{A}_{\Phi_{jj}}=1$.

\begin{equation}\label{eq28}
\left[ \begin{array}{c} \textbf{W}_{ij} \\ \textbf{W}_{ii} \\ \textbf{W}_{kk} \\ \textbf{W}_{kl}\end{array} \right] = \left[ \begin{array}{c} \textbf{W}_{ji}=\alpha  \\ \textbf{W}_{jj} = 1-\alpha \\ 1, \ \text{if} \ k\neq ij \\ \text{on all other edges} \end{array} \right]
\end{equation}
\noindent
where $\alpha=\frac{1}{2}$ and $\beta = \frac{1}{n_i}$
The matrix $\mathbb{E}[\textbf{A}_{\Phi_{ij}}]$ is however more similar to that presented for neighborhood gossip algorithm \cite{benezit2009distributed} except $\textbf{A}_{\Phi_{ii}}=1$. Since the scheduled nature of our algorithm ensures that nodes wake up uniformly in sequential form and choose a neighbor uniformly at random based on neighborhood topology, then,
\begin{equation}
    \mathbb{E}[\textbf{A}_{\Phi_{ij}}] = \frac{1}{N}\sum_{j\in n_i}^{} \frac{1}{|n_j|}, \ i\neq j
\end{equation}

\noindent
where $\mathbb{E}[.]$ indicate an expectation (mean) operator.
It is also stated in \cite{denantes2008distributed} that convergence is sufficiently and necessarily ensured if 
$\lambda_{2}(\mathbb{E}[\textbf{A}_{\Phi_{ij}}])<1$. However, this also depends on whether $\mathbb{E}[\textbf{A}_{\Phi_{ij}}]$ is doubly stochastic. In any case, the proves of these two conditions depends on the nature of the graph. In \cite{denantes2008distributed}, the convergence of both pairwise and neighborhood gossip algorithms are proved for random graphs. Since we have proved the the proximity in entries and structure of $\textbf{A}_{\Phi}$ to $\textbf{W}$ for both algorithms, it suffices to assume that our algorithm with also converge to consensus. A more rigorous proofs of convergence to consensus and convergence times and speeds for different graphs will be presented in later studies. Convergence of states to consensus in further investigated through numerical simulations.
\section{Numerical Simulations}
In this section, we present a numerical evaluation of our algorithm. To measure the performance of the algorithm using simulations, first we define an error measure named the \textit{drift} from the mean, which is defined as:
\begin{equation}
d(k) :=  \Big|\frac{1}{|\textbf{v}|}\sum_{l}^{} x_l(k) - x_{avg}\Big|
\end{equation}

This parameter gives a general indication of the deviation of the states values of each node from the expected mean, which is reminiscent of asymptotic variance. In addition to this metric, we define another error parameter, called neighborhood consensus error or \textit{disagreement}, denoted as $\epsilon$ that shows the performance of the system as the number of nodes increase. This parameter is calculated as:

\begin{equation}
    \epsilon(k) :=  \sqrt{\frac{1}{|\textbf{v}|} \sum_{i,j}^{} \textbf{A}_{ij} \Big[x_{i}(k) - x_{j}(k)\Big]^2}
\end{equation}

\begin{figure}[!p] 
	\centering    
	\includegraphics[width=0.450\textwidth]{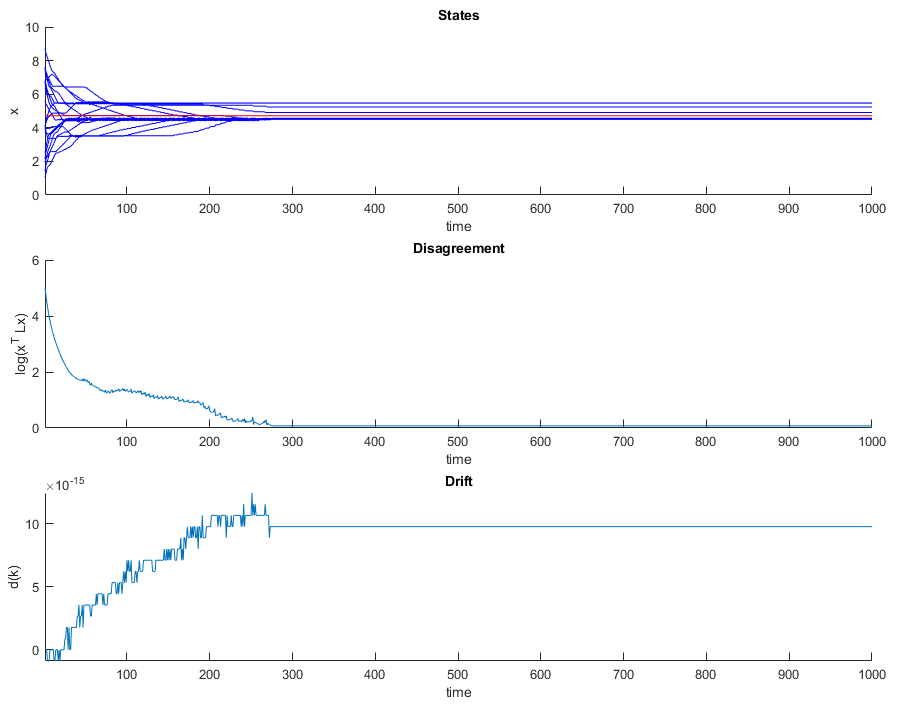}
	\caption[Performance of an Undirected Circular Graph]{Performance of an Circular Graph}
	\label{fig:circular_undirected_50}
\end{figure}

\begin{figure}[!p] 
	\centering    
	\includegraphics[width=0.450\textwidth]{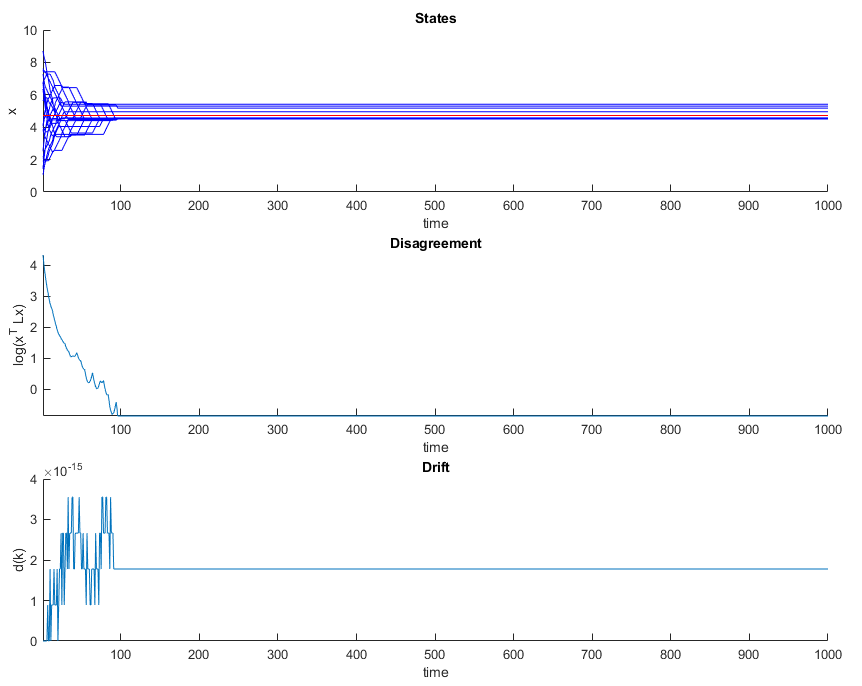}
	\caption[Performance of a Directed Circular Graph]{Performance of a Directed Circular Graph}
	\label{fig:circular_directed_50}
\end{figure}

\begin{figure}[!p] 
	\centering    
	\includegraphics[width=0.450\textwidth]{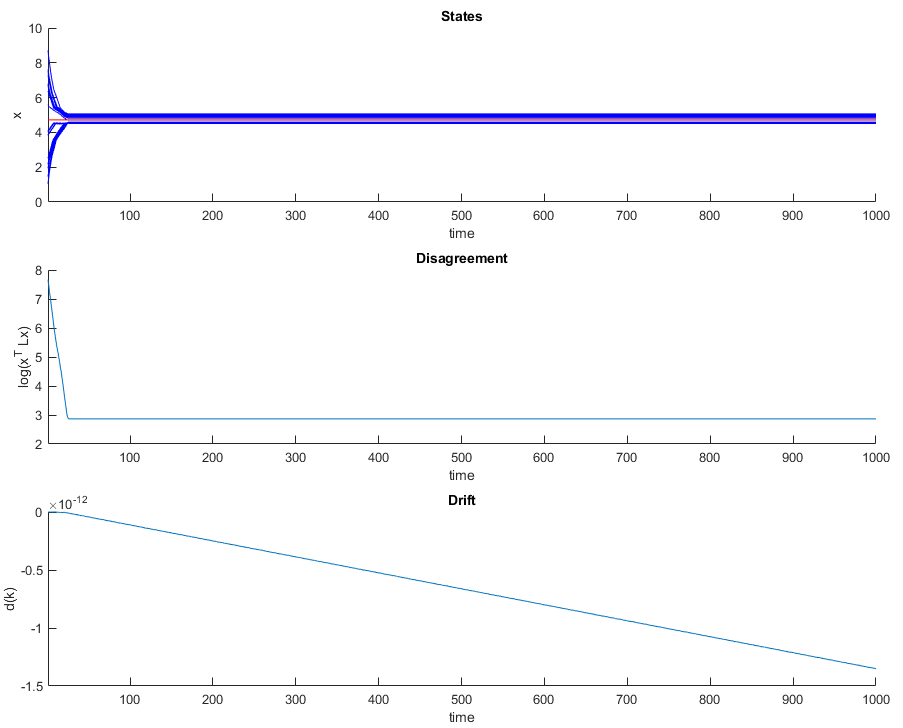}
	\caption[Performance of an Undirected Random Graph]{Performance of an Undirected Random Graph}
	\label{fig:complete_50}
\end{figure}

\begin{figure}[!] 
	\centering    
	\includegraphics[width=0.450\textwidth]{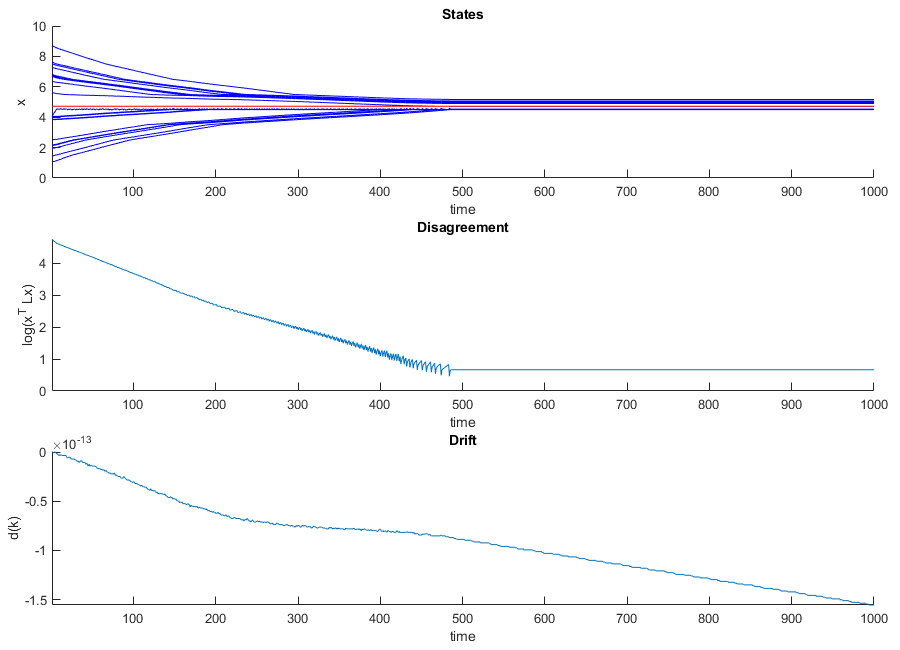}
	\caption[Performance of an Undirected Star Graph]{Performance of an Undirected Star Graph}
	\label{fig:star_50}
\end{figure}

\begin{figure}[H] 
	\centering    
	\includegraphics[width=0.450\textwidth]{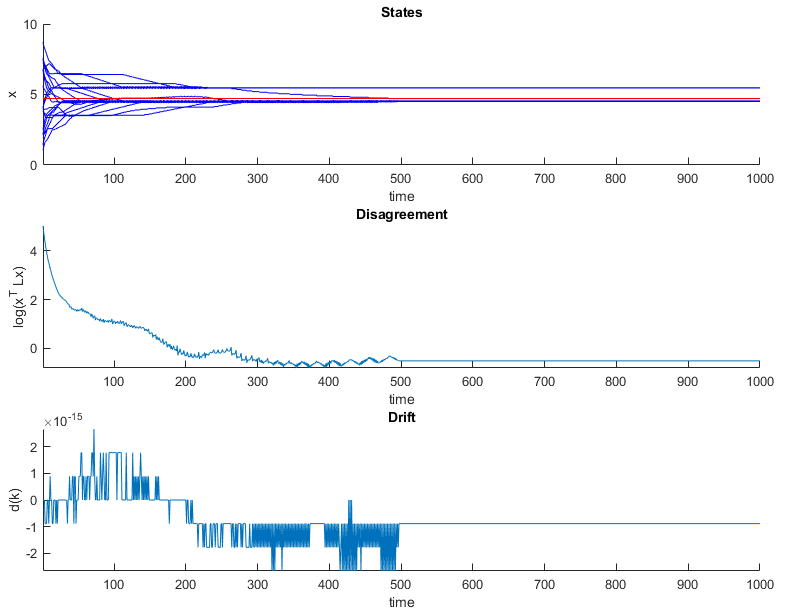}
	\caption[Performance of an Undirected Star Graph]{Performance of an Undirected Chain Graph}
	\label{fig:chain_50}
\end{figure}

We also look at evolution of the states of nodes within the graph. We run simulations on  graphs of 50 nodes in   circular (directed and undirected), random, star and chain topologies as shown in Figures \ref{fig:complete_50},  \ref{fig:circular_directed_50}, \ref{fig:circular_undirected_50}, \ref{fig:star_50}, and \ref{fig:chain_50} respectively. We observe a fine convergence in states for the random graph as compared to the other graphs. The desired average, $x_{ave}$ is shown by a \textquoteleft red line\textquoteright.
Further, we observe drift values below $10^{-15}$ for all topologies. Further more we notice a general decay of disagreement towards zero for all networks. Consensus id observed to be achieved within some margin of deviation for all networks. General convergence time was less than 400 iterations for all networks, with a minimum of 20 iteration for the undirected random graph.

\section{Conclusion}
In this work, we presented a method of achieving asynchronous average consensus for wireless sensor networks which conserves node energy through a randomized controlled duty cycling of nodes. We formulated the asynchronous algorithm in form of a linear switched control system. The convergence and stability of the proposed algorithm was analyzed. Numerical simulations were carried on different graphs to evaluate the proposed algorithm. Numerical results show convergence and very low drift values below others of $10^{-15}$. Future work may focus on effects of quantization noise on the the proposed algorithm. A more thorough convergence analysis (speed and time) in mean sense and in probability could be researched. 

\ifCLASSOPTIONcaptionsoff
  \newpage
\fi

\bibliographystyle{ieeetr}
\bibliography{IEEEabrv,ref}

\end{document}